\documentclass{article} 
\begin{document}

%\twocolumn[\hsize\textwidth\columnwidth\hsize\csname@twocolumnfalse\endcsname
%\pagestyle{plain}

\title{On the dynamics of Kac $p$-spin glasses}

\author{Silvio Franz\\
\it \small The Abdus Salam
ICTP\\
\it 
\small
Strada Costiera 11, P.O. Box 586, I-34100 Trieste, Italy\\ 
\it 
\small
and \\
\it 
\small
Isaac Newton Institute for Mathematical Sciences\\
\it 
\small
20 Clarkson Road, Cambridge, CB3 0EH, U.K.
}

\date{\today}

\maketitle

\begin{abstract}
  This paper discusses the dynamical properties of $p$-spin models
  with Kac kind interactions. For large but finite interaction range
  $R$ one finds two different time scales for relaxation. A first
  relaxation roughly independent of $R$ where the system is confined
  to limited regions of the configuration space and an $R$ dependent
  time scale where the system is able to escape the confining regions.
  I will argue that the $R$ independent time scales can be described
  through dynamical mean field theory, while non-perturbative new
  techniques have to be used to deal with the $R$ dependent scales. 

\end{abstract}

%\pacs{05.20.-y, 75.10Nr}

%\twocolumn\vskip.5pc]\narrowtext

\section{Introduction}
Mean-Field spin glass model, describe glassy phenomena as sharp
ergodicity breaking transitions \cite{MPV}.  It is well known that in
that context two basic route to glassiness are found. In the SK and
similar models, one finds second order phase transitions with a
diverging susceptibility and a continuous order parameter \cite{SK}.
In $p$-spin models, one finds a transition, of the second order kind
from a thermodynamic point of view displays a first order jump in the
order parameter \cite{Gardner}.  While the SK model was proposed as
possible starting points to understand the physics of spin glass
materials, $p$-spin like models have deserved a lot of attention as
mean-field schematization to understand the structural glass
transition of supercooled liquids \cite{W1}.  In particular, the
physical analysis of the model shows the presence of a phase where
despite no sign of thermodynamic singularity partition function is
dominated by an exponential number of metastable states, while the
off-equilibrium dynamics fails to reach equilibrium and falls into an
asymptotic aging states with internal energy extensively higher then
the equilibrium one \cite{leticia}. The relaxation time to equilibrium
is infinite at all temperature below the dynamical transition. As the
Cortona meeting has testified, progresses have been recently achieved
in the comprehension of both the thermodynamics and the dynamics of
Mean-Field models. On one hand, the Guerra \cite{G} and Talagrand
\cite{T} analysis shows that Parisi ansatz, which in these models take
the simple one step form, describes well the thermodynamics, on the
other Ben Arous, Dembo and Guionnet \cite{BaDG} could prove that the
physicist's equations -and underlying assumptions leading to their
derivations- can be fully mathematically justified.

The situation is much less clear as one wants to deal with systems
with finite range interactions. The low temperature properties of
finite range spin-glass models, despite remarkable attempts of
mathematical analysis \cite{NS} remain an open issue of scientific
discussion. While there is no consensus about the fate of mean-field
thermodynamical spin glass phases when the range of interaction is
finite, everybody would agree that the metastability phenomena
associated to the dynamical transition in $p$-spin models should
become cross-overs as soon as the range of the interaction takes a
finite range character.  Possible scenarios for this cross-over have
been put at the basis of a phenomenological theory of the glass
transition known under the name of 'random first order transition
scenario' which describes activated processes responsible for
ergodicity restorations in terms of an effective droplet model where a
bulk restoring force proportional to the configurational entropy
competes with interface free-energy terms that oppose to relaxation
\cite{W1,W2,P1,BB}. Computations of the resulting effective barrier in
the context of microscopic models have been presented in \cite{io,D0}.
Many researchers believe that an accomplished microscopic theory of
this cross-over could provide important hints about the relaxation
processes in supercooled liquids and the glass transition \cite{gold}.

In order to understand the relation between mean field and finite
range systems one can use the classical tools of Kac models
\cite{Kac}, where the interaction range $R$ is considered a tunable
parameter. One would like first to understand the equilibrium and
dynamics in the Kac limit where the interaction range is sent to
infinity after the thermodynamic limit \cite{Kac,lp}, and then to
study the dynamics for finite $R$ and metastability effects in an
asymptotic expansion in $R$. Recently a full mathematical control of
the Kac limit for thermodynamics has allowed to prove convergence of
free-energy and local correlations to the the mean-field values in
p-spin models for even $p$ \cite{FT1,FT3}. In this paper I would like
to extend the analysis to the dynamics. In particular I will consider
the case of the $p$-spin model and study: (a) the dynamics in the Kac
limit (b) the form that the relaxation time should take in an
asymptotic expansion in the range of interaction.  This will lead to a
picture where the long time relaxation can be seen as a passage from
metastable state to metastable state, and the relaxation time computed
as a decay rate.  The analysis I will present will rely strongly on
heuristic derivations. I believe that the analysis of the Kac limit,
clear from the physical point of view, can be made fully rigorous with
mild adaptation of the techniques of \cite{BaDG}. Conversely, the
dynamics on $R$ dependent time scales, remains to a large extent a
challenge for theoretical physics, and the mathematical analysis seems
to me far away in time.

In several occasions in this paper I will refer to the dynamical and
static properties of the mean-field $p$-spin model as they are known
from the physicist's analysis. For many of these properties
mathematical proofs are lacking. Throughout my reasoning I will ignore
this fact and assume that the mentioned properties are correct.

Some of the results presented in this paper have appeared in 
form in \cite{F-epl} and \cite{io}.

\section{The Kac Spherical $p$-spin model}

In order to study the relaxation in a Langevin setting, I use a
variation of the locally spherical version of the Kac p-spin model
considered in \cite{FT1}. Consider $\Lambda_L=\{1,2,...,L\}^d$ the
hypercubic lattice of size $L$ and periodic boundary conditions,
partitioned into hypercubic boxes $B_x$ of size $\ell$ ($L/\ell$ is a
large integer), $x\in \Lambda_{L/\ell}$. The model is defined in terms
of real variables $S_i\in R$ subject to the local spherical constraint
$|\ell^{-d} \sum_{i\in B_x} S_i^2 -1|< \epsilon $ for some small
epsilon and interacting through the random Hamiltonian
\begin{eqnarray}
H(S)=-\sum_{(i_1,...,i_p) \in \Lambda_L^p} J_{i_1,...,i_p}
S_{i_1}...S_{i_p}
\end{eqnarray} 
where the couplings $J_{i_1,...,i_p}$ are for each $p$-uple of the
lattice i.i.d. normal variables with zero mean and variance 
\begin{eqnarray}
E( J_{i_1,...,i_p}^2)= \sum_{l\in \Lambda_L} \frac{1}{R^{pd}}
\psi(|l-i_1|/R)...\psi(|l-i_p|/R). 
\label{psi}
\end{eqnarray}  
The function $\psi: R^+\to R^+$, is chosen to be positive $\psi(x)\geq
0$ and normalizable, $\int_0^\infty \psi(x) dx =1$. Notice that only
variables that are at distances of order $R$ or lower can effectively
interact. The usual mean-field model is recovered choosing $R=L$ and
$\psi(x)=1_{x\leq 1}$.  I will consider instead the following regimes:
(a) the Kac limit where $L\to\infty$, $R\to\infty$, $\ell\to\infty$ in
the specified order, (b) the regime 
\begin{eqnarray}
L>>R>>\ell>>\log(L)
\label{hier}
\end{eqnarray}
where non trivial finite dimensional effects can be expected to take
place \cite{lp2}. 
Consider the Langevin dynamics
\begin{eqnarray}
\frac{d S_i(t)}{dt}= -\mu_x(t) S_i(t) +
\sum_{(i_2,...,i_p)\in \Lambda_L^{p-1}} J_{i,i_2,...,i_p} 
S_{i_2}(t)...S_{i_p}(t)
+h_i(t) +\eta_i(t)
\end{eqnarray}
where the $\mu_x(t)$ form a set of Lagrange multipliers chosen in such
a way to enforce the spherical constraint in average at each time:
$E\langle \ell^{-d} \sum_{i\in B_x} S_i(t)^2\rangle=1$, and
$\eta_i(t)$ is white noise with amplitude $\sqrt{2T}$ and $h_i$ is an
arbitrary extra field term that will be used to generate response
functions. The equation will be interpreted in the Ito sense, and will
be thought to be regularized at small times (e.g. discretized) so that
a regular solution -differentiable with respect to $h$- will exist at
all times.

Below the temperature $T_d=\sqrt{(p(p-2)^{p-2}/(p-1)^{p-1})}$ the mean
field model is in a non ergodic regime.  One can expect for finite $R$ the
existence of two kind of time scales: time scales independent of $R$,
which should coincide with the ones of the corresponding mean field
model and are ineffective to relax at low temperature, and $R$
dependent time scales on which there is ergodicity restoration.  In
order to understand the dynamics of the systems, it is interesting to
consider its evolution in the Kac limit, where, in the order,
$L\to\infty$, $R\to\infty$, $\ell\to\infty$ for fixed value of time.
I would like to argue that for uniform initial conditions, the local
correlation and response functions, defined respectively as
\begin{eqnarray}
C^{L,R,\ell}_x(t,u)=\frac{1}{\ell^d} \sum_{i\in B_x} S_i(t)S_i(u)\nonumber\\
R^{L,R,\ell}_x(t,u)=\frac{1}{2 T \ell^d} \sum_{i\in B_x}
S_i(t)\eta_i(u)
\label{CR}
\end{eqnarray}
tend with probability one to non-random functions $C(t,u)$ $R(t,u)$ of
the time arguments $t$ and $u$, independent of $x$ that verify the
usual equations describing the dynamics of the mean field model (see
e.g. \cite{leticia} for their detailed form and a review of mean-field
off-equilibrium dynamics). 

Unfortunately, I cannot at the moment offer a proof of this
statement. I would like however to discuss the indications that this
is the case. The derivation parallels derivation in the
usual mean field case, and shows that large $R$ in the Kac case is
analogous to large $N$ in the mean field case. 

\subsection{An exact equation}
In this section I derive an exact equation, valid for all values of
the parameters and all kind of initial conditions, that is a
consequence of the Gaussian distribution of the couplings.
Consider observables $A[S,\eta]$ depending on the values of the
variables and the Brownian noise at different times. 
Choices of interest will be $A[S,\eta]=S_i(u)$ and
$A[S,\eta]=\eta_i(u)$. I will denote with angular brackets
$\langle\cdot\rangle$ the average over the initial conditions and the
thermal noise, and with $E(\cdot)$ the average over the quenched
couplings $J_{i_1,...,i_p}$. From the Langevin equation, using partial
integration on the Gaussian couplings one gets: 
\begin{eqnarray}
& &E \langle \frac{d S_i(t)}{dt} A[S,\eta] \rangle = -E \langle
  \mu_x(t) S_i(t)
  A[S,\eta]\rangle
+
\langle  A[S,\eta] \eta_{i}(t)
\rangle +
\nonumber 
\\ & &
\sum_{(i_2,...,i_p) \in \Lambda_L^{p-1}} E (J_{i,i_2,...,i_p})^2
  E \frac{\partial }{\partial J_{i,i_2,...,i_p} }
\langle  A[S,\eta] S_{i_2}(t)...S_{i_p}(t)
\rangle
\end{eqnarray}
In order to make explicit the $J$ derivative, one can use the
Martin-Siggia-Rose, Girsanov representation of the joint probability
of paths $S_i(t)$ and Brownian noise $\eta_i(t)$ starting from an
initial condition $S_i(0)$ chosen with probability $\mu(S(0))$
\begin{eqnarray}
P[S,\eta]\mu(S(0))= \int_{-\infty}^{\infty}
 \prod_{u=0}^t \frac{d\hat{S}_i(u) }{2\pi}
\exp\left(\sum_{i\in\Lambda} \int_0^t du\;  i\hat{S}_i(u) \left[
\dot{S}_i(u)+
\mu_x(u)S_i(u)+
\right.
\right.
\nonumber\\
\left.
\left.
\sum_{(i_2,...,i_p)}
J_{i,i_2,...,i_p}S_{i_2}(u)...S_{i_p}(u) +h_i(u)
+\eta_i(t)
\right] -\frac{1}{2T}\eta_i^2(t)
\right)
\mu(S(0))
\end{eqnarray}
and observe, through integration by part, that insertion of $
i\hat{S}_i(u) $ in correlation functions of the kind $\langle
B[S,\eta]\rangle$, which improperly will be denoted as $\langle
i\hat{S}_i(u) B[S,\eta] \rangle$, acts as differentiation with respect
to $h_i(u)$, or in turn as insertion of $\frac{1}{2T} \eta_i(u)$
\begin{eqnarray}
\langle i\hat{S}_i(u) B[S,\eta] \rangle=\frac{\partial}{\partial
  h_i(u)} \langle B[S,\eta] \rangle =\frac{1}{2T} \langle B[S,\eta]\eta_i(u) \rangle 
\end{eqnarray}
so that, 
\begin{eqnarray}
&& E \frac{\partial }{\partial J_{i,i_2,...,i_p} }
\langle  A[S,\eta] S_{i_2}(t)...S_{i_p}(t)
\rangle =\nonumber\\
&& \int_0^t dv\;  E \left( \frac{1}{2T}\langle 
\eta_i(v)S_{i_2}(v)...S_{i_p}(v)
 A[S,\eta] S_{i_2}(t)...S_{i_p}(t)
\rangle
\right.\nonumber\\
&& \left.
+\sum_{l=2}^p \langle 
S_i(v)S_{i_2}(v)...\eta_{i_l}(v)...S_{i_p}(v)
 A[S,\eta] S_{i_2}(t)...S_{i_p}(t)
\rangle
\right)\nonumber\\
&&+
 E \left( \langle 
\frac{\partial \log(\mu(S(0)))}{\partial J_{i,i_2,...,i_p}}
 A[S,\eta] S_{i_2}(t)...S_{i_p}(t)
\rangle\right).
\label{10} 
\end{eqnarray}
\subsection{The off-equilibrium case}
If one choose an initial distribution independent of the quenched
coupling, so that $\frac{\partial \log(\mu(S(0)))}{\partial
  J_{i,i_2,...,i_p}}=0$, the last term in eq. (\ref{10}) is zero. We
assume in this section that this is the case, and that in addition
$\mu$ is a translation invariant measure. One can consider for
instance the uniform measure on the set ${\cal I}=\{S\in R^{L^d}:\;
|\ell^{-1}\sum_{i\in B_x} S_i^2-1|<\epsilon\; \forall x\}$.

Defining the empirical
correlation function and response: 
\begin{eqnarray}
&&\hat{C}_l(t,u)=\frac{1}{R^d}\sum_i \psi(|l-i|/R) S_i(t) S_i(u)
\nonumber\\
&&\hat{R}_l(t,u)=\frac{1}{2 T R^d}\sum_i \psi(|l-i|/R) S_i(t)
\eta_i(u)
\label{empir}
\end{eqnarray}
and 
\begin{eqnarray}
& &E \langle \dot{ S}_i(t)A[S,\eta] \rangle = -E \langle \mu_x(t)  S_i(t)
  A[S,\eta]
\rangle
 +
E\langle  A[S,\eta] \eta_{i}(t)
\rangle
+\nonumber
\\ & &
\frac{1}{R^d}\sum_l \psi(|l-i|/R) 
p \int_0^t dv\; E\left(
\frac{1}{2 T} \langle 
\eta_i(v) A[S,\eta] \hat{C}_l(t,v)^{p-1}
\rangle
\right.
\nonumber\\ &&
\left.
+(p-1) \langle  S_i(v) A[S,\eta] \hat{R}_l(t,v)\hat{C}_l(t,v)^{p-2}
\rangle
\right) 
\end{eqnarray}
This is an exact equation valid for all values of $L$, $R$, and
$\ell$.  From this equation one sees that if for typical realization
of thermal noise and initial conditions, in the Kac limit the
empirical functions $\hat{C}_l(t,u)$ and $\hat{R}_l(t,u)$ are
self-averaging and tend to space homogeneous ($l$-independent)
limiting functions $C(t,u)$ and $R(t,u)$, then the Lagrange
multipliers would be $x$ independent and one would have:
\begin{eqnarray}
& &E \langle \dot{ S}_i(t)A[S,\eta] \rangle = -\mu(t)E \langle S_i(t)
  A[S,\eta]
\rangle
 +E \langle  A[S,\eta] \eta_{i}(t)
\rangle
+
\nonumber \\ & &
p \int_0^t dv\; E\left( 
\frac{1}{2 T} \langle 
\eta_i(v) A[S,\eta] 
\rangle
{C}(t,v)^{p-1}
\right.\\
\nonumber
&&
\left. 
+(p-1) \langle S_i(v) A[S,\eta]\rangle R(t,v)C(t,v)^{p-2}\right). 
\label{exact}
\end{eqnarray}
where I have used $\frac{1}{R^d}\sum_l\psi(|l-i|/R)\to 1$.  The being
valid for any $A$, it implies, in law, the effective, single site
Langevin equation:
\begin{eqnarray}
\dot{S}(t)=-\mu(t)S(t)+\eta(t)+\xi(t)+(p-1) \int_0^t dv\;
C(t,v)^{p-2}R(t,v) S(v)
\label{single}
\end{eqnarray} 
where $\xi(t)$ is a Gaussian field with correlations $\overline{(
\xi(t)\xi(u))}= p C(t,u)^{p-1}$, and subject to the consistency
conditions: 
\begin{eqnarray}
&&C(t,u)=\overline{(S(t)S(u))}\nonumber\\
&&R(t,u)=\frac{1}{2 T} \overline{(S(t)\eta(u))}.
\end{eqnarray} 
where the bar indicates average over the distribution of the process. 
This process coincides with the one found in mean field dynamics and 
predicts that while at high temperature the system equilibrates after
a finite time, at low temperatures $T<T_d$ the system fails to
equilibrate and enters an off-equilibrium aging regime where the
energy remains extensively higher then the equilibrium one. 

As soon as one gets away from the Kac limit and the range of
interaction is finite, the relaxation time to equilibrium should
always be finite. What the Kac limit is telling to us, is that for
large enough $R$ the relaxation time is essentially $R$ independent at
high temperature, while it becomes $R$ dependent for temperatures $T
{{<} \atop {\sim}} T_c$.

A proof of self-averaging and existence of the limit for the empirical
correlation and response can be envisaged according to the lines of
\cite{BaDG} for the mean field case. The main formal difference with
respect to that case is that the white average over sites that is
responsible for the self-averaging properties of the correlation and
response functions in mean-field is here substituted by the weighed
average (\ref{empir}). The self-averaging property, with existence of
the limit, should appear when the number of variables participating to the
average diverges. Space homogeneity follows from the hypothesis of
statistical homogeneity of the initial condition. 

For the analysis that will follow it is interesting to investigate the
behavior of the fluctuations for finite $L$, $R$ and $\ell$.  of
correlation and response on the scale $\ell$ of the boxes $B_x$.
Equation (\ref{single}) suggests that for fixed times $t,s$, typical
fluctuations of correlations and responses (\ref{CR}) are of order
$O(\ell^{-d/2})$. Assuming the existence of a finite correlation
length beyond which fluctuations become effectively independent,
extreme value statistics implies that the maximum fluctuation is of
order $O(\sqrt{\ell^d/\log(L)})$. The hierarchy (\ref{hier}) between
length scales guarantees that with appropriate choice of 
$L,R,\ell$, this maximal fluctuation can be made arbitrarily
small.

\subsection{Equilibrium dynamics}
To study the largest time scales in the system we need to understand
equilibrium dynamics. As in the previous section, let us consider the
Kac limit. The main difference with the off-equilibrium case is that
the  distribution of the initial state
\begin{eqnarray}
\mu(S_0)=\frac{1}{Z} e^{-\beta H(S_0)}
\end{eqnarray}
now depends on the quenched couplings. In the derivation to the
analogous of equation (\ref{exact}) one should keep into account that 
\begin{eqnarray}
\frac{\partial \mu(S)}{\partial J_{i,i_2,...,i_p}}=S_{i_1}...S_{i_p}-
\langle S_{i_1}...S_{i_p}\rangle_{eq}
\end{eqnarray}
this gives rise to an additional term into the exact equation
(\ref{exact}) for $E \langle \dot{ S}_i(t)A[S,\eta] \rangle $, that
then reads:
\begin{eqnarray}
& &E \langle \dot{ S}_i(t)A[S,\eta] \rangle = -\mu_x(t)E \langle S_i(t)
  A[S,\eta]
\rangle
 +E\langle  A[S,\eta] \eta_{i}(t)
\rangle
+
\nonumber
\\ & &
\frac{1}{R^d}\sum_l \psi(|l-i|/R) 
p \int_0^t dv\; E
\left(
\frac{1}{2 T} \langle 
\eta_i(v) A[S,\eta] \hat{C}_l(t,v)^{p-1}
\rangle
+
\right. 
\nonumber
\\
&&
\left. 
(p-1) \langle  S_i(v) A[S,\eta] \hat{R}_l(t,v)\hat{C}_l(t,v)^{p-2}
\rangle\right) +\nonumber\\ &&
\frac{1}{R^d}\sum_l \psi(|l-i|/R) 
E \langle 
S_i(0) A[S,\eta] \hat{C}_l(t,0)^{p-1}
\rangle
\nonumber
\\
&&
-\sum_{l,i_2,...,i_p} \frac{1}{R^{pd}}\sum_l \psi(|l-i|/R)
  \prod_{r=2}^p\psi(|l-i_r|/R)  \times\nonumber\\ &&
E\left( \langle 
 A[S,\eta] S_{i_2}(t)...S_{i_p}(t)
\rangle
\langle 
 S_i S_{i_2}...S_{i_p}
\rangle_{eq}\right)
\label{eq}
\end{eqnarray}
In order to write closed equations in this case, besides hypothesizing
the self-averaging condition for $\hat{C}_l$ and $\hat{R}_l$ discussed
in the previous section, one needs to know the structure of the
equilibrium correlations that appear in (\ref{eq}). The results of
\cite{FT1} imply that the static correlations on scales of order $R$,
tend to the corresponding mean field function in the limit. I am
interested here to the regime $T_k<T<T_d$ where these static
correlations are vanishingly small in the limit, so that the last term
in (\ref{eq}) will be zero. In equilibrium conditions the fluctuation
dissipation theorem and time translation invariance holds and
$R(t,s)=R(t-s)=\beta \frac{\partial C(t-s)}{\partial s}$. The
resulting equation for $C$ reduces after a little algebra to the usual
equilibrium mean field equation \cite{Abarrat}:
\begin{eqnarray}
\frac{d C(t)}{dt}=-T C(t)+\frac{p}{2T} \int_0^t ds\;
C^{p-1}(t-s)\frac{d C(s)}{ds}
\label{mct} 
\end{eqnarray}
The behavior of this equation is very well known. Above $T_d$ the
correlation function decays in a finite time to its equilibrium vale
$\lim_{t\to\infty} C(t)=0$, while below $T_d$, ergodicity is broken
and $\lim_{t\to\infty} C(t)=q_{EA}$.  $q_{EA}$ is the Edwards-Anderson
parameter, which does not coincide with the static value of the
correlations in the Kac limit
$\lim_{R\to\infty}\frac{1}{\ell^d}\sum_{i\in B_x}E\langle
S_i\rangle_{equil}^2=0$. 

We are in a situation where the Kac limit does not commute with the
large time limit.  A system with a finite but large $R$, for a large
time will follow closely the evolution dictated by (\ref{mct}), but,
eventually, will be able to relax below $q_{EA}$ on $R$ dependent time
scales.  Increasing $R$, one can make the relaxation time scale as
long as wanted, in particular, it can be chosen so long to let the
time to the system to explore ergodically the region where the local
correlations $C_x(t)$ never go below a certain value $q_0<q_{EA}$.
This is a crucial observation that allows to identify typical
equilibrium configurations as belonging to metastable states as it will
be discussed in the next section. 

\section{$R$ dependent time scales} 

On time scales diverging with the interaction range, the independence
between different spins which is implied by the self-averaging
property of the correlation function should fail and the mean-field
dynamical equations loose their validity.  In order to investigate the
relaxation below $q_{EA}$ one should proceed in a different manner,
reminiscent of the Lebowitz and Penrose \cite{lp2} analysis of
metastability in Kac models of first order phase transition.

I said that is a consequence of the fact that the ergodic time is
divergent with $R$, for $R$ large enough the equilibrium dynamics has
the time to explore ergodically the region where for all boxes $x$ the
correlation $C_x(t)>q_0$.  This condition defines this region as
metastable state in a sense that extends the one first specified in
\cite{lp2} in connection with Kac models with first order phase
transitions. In fact one has the following properties: 1) The system
finds itself in states of constrained equilibrium. 2) The time to
leave these states is large, but much larger is the time to come back
once departed. 3) The correlation function is homogeneous in space.
Formally, for typical equilibrium configurations $S_i^0$, one can
define the set ${\cal R}(S_0)=\{S\in {\cal I}|\;\; q_x(S,S^0)>q^0\}$
and the metastable state as the constrained equilibrium state
\begin{eqnarray}
\mu_{{\cal R}(S_0)}(S)=\frac{1}{Z_{{\cal R}(S_0)}} e^{-\beta H(S)}1_{{\cal R}(S_0)}(S). 
\end{eqnarray}
Notice that the states defined in this way do not form a partition of
the equilibrium manifold. In fact couples of states so defined,
corresponding to two different reference equilibrium configurations
can overlap if these configurations are close enough to each other.
Despite this fact, one can profitably describe the relaxation as a
passage from metastable state to metastable state, and estimate the
relaxation time as the inverse decay rate of these states. Denoting as
${\cal H}$ the Fokker-Plack operator and as $\overline{\cal R}(S_0)$ the
set complementary to ${\cal R}(S_0)$, the decay rate of ${\cal R}(S_0)$ is:
\begin{eqnarray}
\lambda=
\int_{S\in {\cal R}(S_0); \;\;\; S' \in \overline{\cal R}(S_0)} dS\; dS'\; 
\langle S'| {\cal H} |S\rangle \times \mu_{{\cal R}(S_0)}(S). 
\end{eqnarray}
This expression can be bounded from above noticing that the
Fokker-Planck operator is local, so that if $S'$ has to be in 
$\overline{\cal R}(S_0)$ then $S$ has to be on the border of ${\cal R}(S_0)$, 
\begin{eqnarray}
{\partial \cal
  R}(S_0)=\{S\in {\cal R}(S_0)|\;\; \exists x_0:\;  q_{x_0}(S,S^0)=q^0\}. 
\end{eqnarray}
Moreover, for any $S$, the integral $\int d S'\; \langle S'| {\cal H}
|S\rangle $ can be bound by a constant $C$ and one gets the estimate
\begin{eqnarray}
\lambda\leq
C \int_{S\in {\partial \cal R}(S_0)} dS\; 
 \mu_{{\cal R}(S_0)}(S)=C \frac{Z_{\partial {\cal R}(S_0)}}{Z_{{\cal R}(S_0)}}
\end{eqnarray}
where $C$ is a constant of order 1.  two comments are here in order:
1) Defined in this way, the escape rate $\lambda$, and its
estimate $\frac{Z_{\partial {\cal R}(S_0)}}{Z_{\cal R}(S_0)}$ are functions of
the reference configuration $S_0$. One can expect however that both
quantities are self-averaging, i.e. assume values independent on $S_0$
with probability approaching 1 in the thermodynamic limit. This
suggests to consider 
\begin{eqnarray} 
\lambda_{typ}\sim\exp\left[E(\log Z_{\partial {\cal R}(S_0)}-\log{Z_{{\cal
      R}(S_0)}})\right]=\exp(-\beta \Delta F(q_0))
\end{eqnarray}
where $E$ stands here for Boltzmann-Gibbs average over the $S_0$ and
quenched average over the Gaussian couplings.  2) The definition of
${\cal R}(S_0)$ and of the decay rate depend on the value $q_0$, which
defines a sort of amplitude of the state. The ``right'' value
optimizing the estimate, is the one such that a local equilibrium
fluctuation of size $q_0$ has equal probability to be reabsorbed in
${\cal R}(S_0)$ and to drive the system out of it. This, together with
the fact that what we get is a lower bound for the decay rate,
suggests that analogously to first order transition kinetics
\cite{L1}, one should maximize $\Delta F(q_0)$ with respect to
$q_0$.

We get a relation between the relaxation time and a free-energy
barrier defined coupling the system with a reference configuration.
On general ground one can expect the barrier to be proportional to the
interaction volume $R^d$. In the next section I will discuss some
theoretical techniques to compute the barrier.

\section{The computation of the barrier}
I would like to discuss here possible strategies for the computation
of the barrier. This is a hard task and my discussion will be highly
conjectural. One needs to compute, for ${\cal S}={\cal R}$ or ${\cal
  S}=\partial {\cal R}$, the following free-energy:
\begin{eqnarray}
F_{\cal S}= \frac{-T}{ Z}
\langle\langle
 \int d S_0 e^{-\beta H(S_0)} \log \int d S
e^{-\beta H(S)} 1_{{\cal R}(S_0)}(S)
\rangle\rangle
\end{eqnarray}
where I denoted with $\langle\langle \cdot \rangle\rangle$ the average
over the quenched couplings $J_{i_1,...,i_p}$.  This free-energy will
consist in an extensive term plus corrections.  In considering the
difference, the extensive term and the corrections due to finite size
effects should compensate in the difference and only the term, finite
in the thermodynamic limit, which is related to the relaxation time
survives.  From the previous analysis, it is clear that $F_{\cal R}$
is dominated by the configurations where in all the boxes indexed by
$x$ one has that $q_x(S,S_0)=q_{EA}$. On the other hand, the
restricted partition function $Z_{\partial {\cal R}(S_0)}$ can be
written as a sum over all sites $x_0$ such that $q_{x_0}(S,S_0)=q_0$,
namely,
\begin{eqnarray} 
&&Z_{\partial {\cal R}(S_0)}=\sum_{x_0} Z_{x_0}\\
&& Z_{x_0}=\int_{S\in {\cal R}(S_0); \;
  q_{x_0}(S,S_0)=q_0} dS e^{-\beta H(S)} 
\end{eqnarray}
One can expect $Z_{x_0}$ to be dominated by a single configuration
consisting in a localized excitation of linear size $O(R)$ around the
point $x_0$. For large $R$, for most sites $x_0$ these excitations
will become self-averaging and independent of the site $x_0$.  The
partition function $Z_{\partial {\cal R}(S_0)}$ will receive
contributions from these sites, but could also receive contributions
from exceptional sites particularly keen to excitation and where the
relaxation would be initiated with the highest probability.  At
present I do not have a clear cut argument to decide if the dominant
contribution to the barrier is given by typical or exceptional sites.
In favor of the first hypothesis one could argue that the condition
$\ell>> \log L$ should guarantee that too exceptional sites are absent
from typical samples. If this is the case, any site $x_0$ is equally
likely to initiate the relaxation independently of $S_0$. In favor of
the second one, one could argue that relaxation below $q_{EA}$
proceeds through amplification of small $O(\ell^{-d/2})$ fluctuations
in the initial condition. This kind of amplification effect has been
observed in realistic short range glass models, where sites that
exhibit faster then average relaxation in the fast regime, are more
likely to initiate the slow part of the relaxation \cite{juanpe}. In that
cases however the interaction range cannot be considered a large
parameter. 

At any rate, the free-energy $F_{\partial{\cal R}(S_0)}$ can be estimated
through the 
the replica method, taking advantage of the identity
\begin{eqnarray}
E(F_{\partial{\cal R}(S_0)})=\lim_{m\to 0\atop n\to 0}-T \langle\langle
 Z^{m-1} \int dS_0
e^{-\beta H(S_0)} (Z_{\partial{\cal
 R}(S_0)(S_0)}^n-1)/n\rangle\rangle. 
\label{rep}  
\end{eqnarray}
As usual, the computation is performed starting from integer values of
$m$ and $n$, and continuing to real values using a an appropriate
modification of Parisi ansatz to take into account the constraints.
It should be possible as in case of the computation of the
unconstrained free-energy, to relate the replica computations to a
variational principle in the space of ``random overlap structures''
\cite{ass}, but in this paper I will not pursue this path.  Instead, I
will sketch the outline of the replica computation of the free-energy
in the simplest possible approximation \cite{FP1,io,F-epl}. For
integer $n$ and $m$ the average over the quenched noise can be
explicitly performed.  One has to introduce local overlaps between
replicas
\begin{eqnarray}
Q_x^{a,b}=\frac{1}{\ell^d} \sum_{i\in B_x} S_i^a S_i^b \;\;\; a,b=1,...,n+m
\end{eqnarray}
where the replicas from 1 to $m-1$ describe the normalization
$Z^{m-1}$, replica number $m$ describes $S_0$ and replicas from $m+1$
to $m+n$ describe configurations in ${\cal R}(S_0)$. In terms of these
order parameters one can write 
the replicated partition function as 
\begin{eqnarray}
    Z_{repl}=\int \prod_{a>m} dx_0^a \int_{Q^{m,a}_x \geq q_0;\;
      Q^{m,a}_{x_0}=q_0\;\; \forall a> m} {\cal D}Q e^{-\frac{1}{T} F[Q]}
\label{FunInt}
\end{eqnarray}
where $F[Q]$
is a coarse grained replica free-energy \cite{FT3}
\begin{equation}
F[Q]=R^d S[Q] =R^d
\int_0^{L/\ell} 
d^dx\; \left[ K( \{ { Q}_{\alpha,\beta}\},x)
 +V(Q(x)) \right]
\end{equation}
with
\begin{eqnarray}
K(  {Q}_{\alpha,\beta},x)&=& \frac{-\beta}{2}\sum_{\alpha,\beta} 
[f ( {\hat Q}_{\alpha,\beta}(x) )-
f({Q}_{\alpha,\beta}(x))]\nonumber\\
V(Q)&=&-
\frac{\beta}{2} \sum_{\alpha,\beta} f({
  Q}_{\alpha,\beta})-\frac{1}{2 \beta}{\rm Tr} \log Q 
\end{eqnarray}
and we have defined 
\begin{equation}
{\hat Q}_{\alpha,\beta}(x)=\int dy\; \psi(x-y)  Q_{\alpha,\beta}(y).
\end{equation}
Notice that thanks to the scaling $F[Q]=R^d S[Q]$ for large $R$ one
can evaluate the integral (\ref{FunInt}) by saddle point.  I would
like here just briefly discuss the results of the simplest ansatz for
the matrix $Q_{ab}(x)$, obtained assuming replica symmetry. It is
clear from (\ref{rep}) that the integer $n$ and $m$ partition function
is invariant under the group of $S^{m-1}\times S^n$ of permutation of
the $m-1$ replicas ($a=1,...,m-1$ in our notation) coming from the
denominator among themselves and the $n$ ($a=m+1,...,n+m$) replicas in
$\partial{\cal R}(S_0)$ among themselves.  Within the replica
symmetric framework one chooses $x_0^a=x_0$ for all $a>m$. This ansatz
for $x_0^a$ is the translation into the replica formalism of the
hypothesis discussed above that independently of $x_0$ all sites are
equally likely to start the relaxation. Replica symmetry also
implies that following structure of the matrix $Q_{ab}(x)$:
\begin{equation}
  Q_{a,b}(x)= \left\{ 
\begin{array}{lll}
& 1   & a=b\\
& s(x) & a\ne b \;\; a, b=1,...,m-1 \\
& p(x) & a<m, \;\; b>m \;\;{\rm or} \;\; a>m, \;\; b<m \\
& q(x) & a=m, \;\; b>m \;\;{\rm or} \;\; a>m, \;\; b=m \\
& r(x) & a\ne b \;\; a,b\ge m
\end{array}
\right.
\end{equation}
It can be seen \cite{FP1} that the equation for $s(x)$ decouples from the other
variables and is solved by $s(x)=p(x)=0$. As far as the other parameter 
are concerned, a further simplification consists in considering saddle
points with $r(x)=q(x)$, which can be verified to exist \cite{io}. 
With this ansatz, the various terms of the free-energy become, in the
limit of small $n$: 
\begin{eqnarray}
&&K(\{q(x)\},x)=-n\frac{\beta}{2} [\hat{q}^p(x)-q^p(x)]
\nonumber\\
&& V(q(x))= -n\left(\frac{\beta}{2} q^p(x)+\frac{1}{2\beta}
  [\log(1-q(x))+q(x)]\right). 
\end{eqnarray}
and the action functional 
\begin{eqnarray}
S=\int dx[K(x)+V(x)] 
\end{eqnarray}
$V(q)$ is a single minimum function at high temperature ($T> T_d$),
while it has two minima below. The free-energy difference between the
minima is well known to represent the configurational entropy per spin
$\Sigma(T)$ in the mean-field model, i.e.  the logarithm of the
multiplicity of ergodic components divided by the volume \cite{FP1}.
Within the simple approximation in which the replica matrix is
parametrized in term of a single space dependent parameter, we are in
presence of Landau like field theory analogous to the one for systems
with a first order transition. The free-energy barrier can the be
estimated through instantonic techniques, looking at the minimum
action solution with boundary condition equal to $q_{EA}$ at infinity
and to $q_0$ in $x_0$ \cite{P1,io}. Close to $T_k$ this results in a
conventional droplet theory where the free-energy barrier results from
an effective droplet model where a bulk free-energy, in this case
proportional to the configurational entropy, competes with a boundary
term, which for the nature of the theory is just a surface term. The
detailed computation of the barrier \cite{io} leads to a Vogel-Fulcher
like expression
\begin{eqnarray}
\Delta F=R^d \frac{C(T)}{\Sigma(T)^{\gamma_d}}
\label{AG}\end{eqnarray}
with $\gamma_d=d-1$ and $C(T)$ a regular function of $T$.  This
computation supports in the context of a microscopic model the
phenomenological ``entropic droplet'' arguments first introduced in
\cite{W1} and recently revived in \cite{BB}.  Formula (\ref{AG}) is
obtained in the replica symmetric approximation and could be changed
by considering better saddle points.  About this result I have mixed
feelings: on the one hand one finds a Adam-Gibbs like relation of
inverse proportionality between configurational entropy and
free-energy barrier, which is a good thing. On the other, in the
relation one finds an exponent $\gamma_d=d-1$ which would predict a
lower critical dimension of 1 for the ideal glass transition and would
be equal to 2 in three D. This is at variance with the expectation,
based on the Adam-Gibbs theory and verified in numerical simulations
of model glasses, of a linear relation between the free-energy barrier
and inverse configurational entropy in three dimensions.  I believe
that while the general form $\Delta F=R^d
\frac{C(T)}{\Sigma(T)^{\gamma_d}}$ only depends on the fact that the
bulk free-energy difference between the minima is the configurational
entropy, should be robust to improvements of the calculation, the
value of $\gamma_d$ could be modified by better approximations. One
can envisage two sort of modifications in the replica computation: 1)
Within the ansatz that $x_0^a$ is independent of $a$, to find better
saddle points for the matrix $ Q_{a,b}(x)$. This has been attempted in
\cite{D0} where an replica broken solution to the instantonic problem
has been found.  Unfortunately, this solution though changing the
function $C(T)$, it does not change the exponent.  It is my opinion
that saddle points in this class cannot modify the exponent.  2) One
can consider more complex configurations where the position $x_0^a$ of
the center of the instanton depends on $a$. Saddle points of this kind
would break the replica symmetry in a vectorial way \cite{DotMez}, and
could lead to a different value of the exponent $\gamma_d$. The search
of new saddle point to the replica variational problem will be matter
of future research.

\section{Conclusions}

In this paper I presented the beginning of the analysis of the
dynamics of the spherical $p$-spin model with Kac kind of
interactions.  I argued in favor of the existence, both in
off-equilibrium and equilibrium dynamics at low temperature, of two
different time regimes for large $R$.  There is a first regime where
dynamical time scales are independent (or quasi-independent) of $R$.
Evidence for this regime is reached observing that in local
correlations and responses should verify the dynamical mean-field
equations of the infinite range model in the Kac limit. I am confident
that the dynamical equations in this regime will be fully
mathematically justified.  Then there is a second dynamical regime, on
time scales diverging with $R$, where the system manages to explore
ergodically the configuration space. The extreme separation of scales
that one has for large $R$ allows the description of the dynamics as a
passage between metastable states that are explored ergodically before
being abandoned. Thanks to that property, the relaxation time can be
estimated as the escape time from typical states and related to a
well-defined free-energy barrier. The computation of this barrier is
unfortunately a highly non-trivial problem. We presented here a simple
attempt based on replica symmetry predicting an ideal glass transition
for the mode in dimension $d>1$ and giving a modified Adam-Gibbs
relation between free-energy barrier and configurational entropy. The
simple theory we have sketched could fail due to several kinds of
replica symmetry breaking effects. While we believe that the general
relation will continue to be valid, the value of the exponent in
modified Adam-Gibbs relation and then the lower critical dimension for
the ideal glass transition could be changed in the ultimate theory.

{\bf Acknowledgments}

This work was supported in part by the
European Community's Human Potential programme under contract
``HPRN-CT-2002-00319 STIPCO''.

\end{document}